\documentclass{mem}
\usepackage{natbib}\usepackage{txfonts}\usepackage{balance}
\usepackage{graphicx}
\usepackage[a4paper]{hyperref}
\idline{00}{000}
\begin{document}

\title{
TeV Active Galactic Nuclei:\\%
Multifrequency Modeling%
}

\subtitle{}

\author{
J.-P. \,Lenain
}

\offprints{J.-P. Lenain}

\institute{
LUTH, Observatoire de Paris, CNRS, Universit\'e Denis Diderot; 5 Place Jules Janssen, 92190 Meudon, France.
\email{jean-philippe.lenain@obspm.fr}
}

\authorrunning{Lenain}

\titlerunning{TeV AGNs: Multifrequency modeling}

\abstract{In the recent years, the new generation of Imaging Atmospheric \v{C}erenkov Telescopes successfully detected very high energy (VHE; $E>100$\,GeV) $\gamma$-ray emission from a growing number of Active Galactic Nuclei (AGNs), mainly belonging to the blazar class.

Among these recent results, we will mainly focus in this work on two recent discoveries made with the H.E.S.S. experiment: the tremendously active state of the blazar PKS\,2155$-$304 observed in July 2006 and the discovery of VHE $\gamma$-rays from the radio galaxy Cen\,A. On the one hand, the observation of very fast variability in PKS\,2155$-$304 challenges the current radiative standard models for TeV blazars. On the other hand, the discovery of Cen\,A as a source of VHE $\gamma$-rays firmly establishes, together with the previous detection of M\,87, radio galaxies as a new class of VHE emitters.
}
\maketitle{}

\section{Introduction}

With the advent of the current generation of Imaging Atmospheric \v{C}erenkov Telescopes (IACTs) such as H.E.S.S.\footnote{\url{http://www.mpi-hd.mpg.de/hfm/HESS/}}, MAGIC\footnote{\url{http://wwwmagic.mppmu.mpg.de/}}, VERITAS\footnote{\url{http://veritas.sao.arizona.edu/}} and CANGAROO\footnote{\url{http://icrhp9.icrr.u-tokyo.ac.jp/}}, our vision of the sky at very high energy (VHE; $E>100$\,GeV) has dramatically changed. In less than ten years, the number of sources detected at VHE has increased from 5 to more than seventy as of writing this manuscript\footnote{see e.g. the TeVCat catalog online at \url{http://tevcat.uchicago.edu/} for an up-to-date list of known VHE sources.}.

\begin{figure*}[t!]
\resizebox{\hsize}{!}{\includegraphics[clip=true]{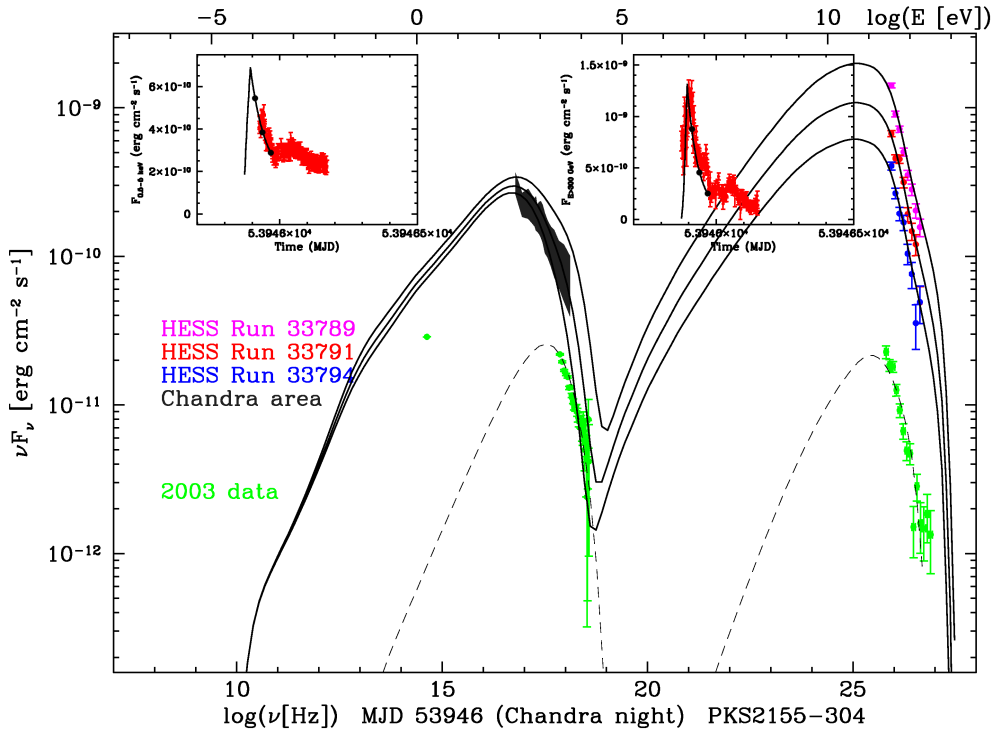}}
\caption{\footnotesize
Spectral energy distribution obtained with the dynamic SSC modeling of the second flare of PKS\,2155$-$304 observed in July 2006 with H.E.S.S. Each solid line corresponds to a snapshot of the evolving SED, simultaneous to the three derived H.E.S.S. spectra during the night. The inlays show the corresponding X-ray ({\em left}) and VHE ({\em right}) light curves, the three spectra represented in the SED corresponding to the three epochs marked in black dots in the light curves.
}
\label{fig-PKS2155_DynamicSED}
\end{figure*}

The majority of them belongs to our Galaxy, while about 25 extragalactic sources have been reported, all of them being active galactic nuclei (AGNs) and most of these being of the blazar class. Blazars are compounded by flat spectrum radio quasars and BL\,Lac, the jet of which is closely aligned to the line of sight, thus amplifying the observed flux by relativistic Doppler boosting. These objects present a double bump-shaped spectral energy distribution (SED), the first bump -- peaking in the optical/X-ray domain -- being generally attributed to leptonic synchrotron emission, while the nature of the second bump -- peaking in the GeV domain -- is more unclear. In leptonic models, it could be attributed to synchrotron self-Compton \citep[SSC, e.g.][]{1992ApJ...397L...5M} or external inverse Compton \citep[e.g.][]{1994ApJ...421..153S} processes.

Hadronic emission could also be responsible for this high energy bump, through the radiation of secondary particles created by proton interaction on protons \citep[e.g.][]{2009arXiv0902.0731O}, photons \citep[e.g.][]{1991A+A...251..723M,2003APh....18..593M} or magnetic fields \citep{2002MNRAS.332..215A}, even though these models can usually hardly interpret fast variability compared to leptonic models, as for example in the case of the fast VHE flux variability observed from 1ES\,1959$+$650 in 2002 \citep{2004ApJ...601..151K}, Mrk\,501 in 2005 \citep{2007ApJ...669..862A}, and from PKS\,2155$-$304 in 2006 \citep{2007ApJ...664L..71A,2008AIPC.1085..415L}.

\section{The multi-blob SSC model}

In a leptonic framework, we developed a multi-blob SSC model \citep{2008A+A...478..111L} to interpret the VHE data taken by H.E.S.S. on the radio galaxy M\,87 \citep{2006Sci...314.1424A}. In this model, an inhomogeneous flow continuously crossing a stationary shock front, located in the innermost part of the broadened jet formation region beyond the Alfv\'en surface as inferred from magnetohydrodynamics models \citep[e.g.][]{2006MNRAS.368.1561M}, could lead to a differential Doppler boosting effect that could lead to a significant blazar-like effect even for sources with a misaligned jet \citep[see][for more details]{2008A+A...478..111L}.

Using this model, we interpreted the VHE spectrum measured by H.E.S.S. in 2004 from M\,87, and predicted possible detectable VHE emission from other misaligned blazar-like sources like Cen\,A, 3C\,273 or PKS\,0521$-$36. Particularly, we predicted a detectable flux from Cen\,A with current IACTs at a $\sim$5$\sigma$ confidence level in $\sim$50\,h of observations.

\section{The blazar PKS\,2155$-$304}

PKS\,2155$-$304 is a blazar known to emit VHE $\gamma$-rays since the early days of VHE $\gamma$-ray astronomy \citep{1999ApJ...513..161C}. In 2003, the H.E.S.S. collaboration reported a low activity state which could be interpreted in both leptonic and hadronic frameworks \citep{2005A+A...442..895A}. However, in July 2006, PKS\,2155$-$304 showed to be in a very active state, and experienced two dramatic flaring events on July 28 \citep{2007ApJ...664L..71A} and July 30, 2006 \citep[][]{2008AIPC.1085..415L}, the second burst exceeding a peak flux of 16 Crab. Moreover, simultaneous X-ray coverage with {\it Chandra} revealed a nearly cubic relationship between the X-ray flux and the VHE flux, ruling out one-zone SSC models for the interpretation of this event.

\begin{figure}[t!]
\resizebox{\hsize}{!}{\includegraphics[clip=true]{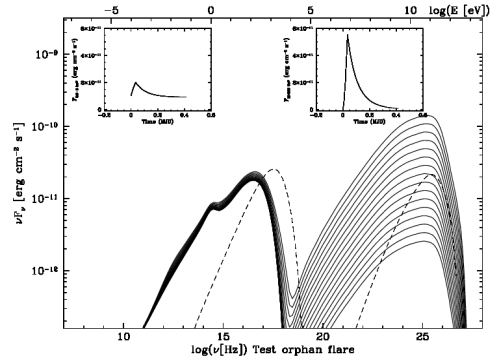}}
\caption{\footnotesize
Dynamic spectral energy distribution for a template solution of an orphan VHE flare event, like the one observed in 1ES\,1959$+$650 in 2002 \citep{2004ApJ...601..151K}. The inlays represent the X-ray ({\em left}) and VHE ({\em right}) light curves. The overall event lasts for $\sim$10\,h. For comparison, the low state of activity of PKS\,2155$-$304 as observed with H.E.S.S. in 2003 is represented in dashed lines.
}
\label{fig-1ES1959_stacked}
\end{figure}

Using a dynamic SSC model \citep{2003A+A...410..101K}, we were able to interpret this rich data set with a two-zone SSC model, in which the synchrotron component of a slowly evolving, extended jet dominates the X-ray radiation, while a small, dense blob dominates the inverse Compton component and is responsible for the fast variability in VHE \citep[see Fig.~\ref{fig-PKS2155_DynamicSED}, and also][]{2008AIPC.1085..415L}. By increasing the density ratio between the extended jet and the blob, we obtained a template solution for an orphan VHE flare event (see Fig.~\ref{fig-1ES1959_stacked}), similar to the one observed in 1ES\,1959$+$650 in 2002 (in this context see also \citet{2008MNRAS.390..371K} for a similar interpretation of the first flare of July~28, 2006).

\section{The radio galaxy Cen\,A}

\begin{figure}[t!]
\resizebox{\hsize}{!}{\includegraphics[clip=true]{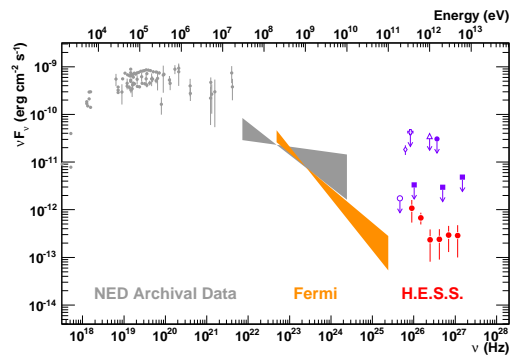}}
\caption{\footnotesize
High energy part of the spectral energy distribution of Cen\,A, with archival data from {\it CGRO}, the recent {\it Fermi}/LAT \citep{2009arXiv0902.1559A} and H.E.S.S. \citep{2009ApJ...695L..40A} spectra.
}
\label{fig-CenA_SED}
\end{figure}

Cen\,A is the nearest radio galaxy, of the FR-I type, at a distance of 3.8\,Mpc. The H.E.S.S. collaboration carried out observations on this object between April 2004 and June 2008, more than 60\% of the data being taken in 2008. After more than 100\,h of observations, Cen\,A was eventually detected in the VHE domain as a point-like source at the 5$\sigma$ confidence level, and the corresponding apparent luminosity was derived to be $L(E>250\,\mathrm{GeV}) \approx 2.6 \times 10^{39}$\,erg\,s$^{-1}$ \citep[see Fig.~\ref{fig-CenA_SED}, and][for more details]{2009ApJ...695L..40A}. This discovery, together with the previous detection of M\,87, firmly establishes radio galaxies as a new class of VHE emitters.

Given the poor spatial resolution in the VHE domain compared to other wavelengths, many different interpretations are consistent with the detection of VHE $\gamma$-rays from Cen\,A, however the giant radio lobes can be excluded as the VHE emitting zone. For example, Cen\,A could be seen as a misaligned blazar \citep[see e.g.][]{2005A+A...432..401G,2008A+A...478..111L}. Hadronic models could also possibly account for these data \citep[see e.g.][for an application to M\,87]{2004A+A...419...89R}. VHE emission could arise from extended structures, by external inverse Compton on the starlight radiation \citep{2006MNRAS.371.1705S}, or in analogy with a supernova remnant-type process at a large scale as suggested by the recent results of \citet{2009MNRAS.395.1999C} conducted with {\it Chandra} and revealing efficient leptonic acceleration in the inner south-western radio lobe. It has also been argued that the immediate vicinity of the central supermassive black hole could be a VHE emitting zone \citep[see][for an application to M\,87]{2007ApJ...671...85N,2008A+A...479L...5R}.

\begin{figure}[t!]
\resizebox{\hsize}{!}{\includegraphics[angle=-90,clip=true]{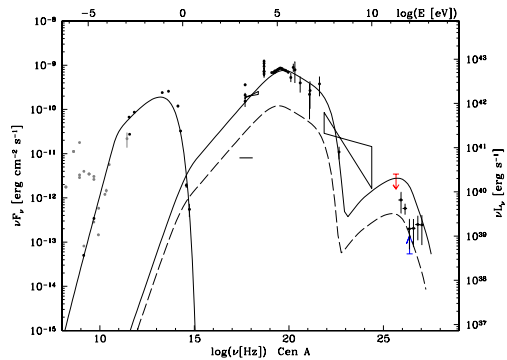}}
\caption{\footnotesize
SED of Cen\,A assuming that the soft $\gamma$-ray emission detected by {\it CGRO} is of synchrotron nature within the multi-blob SSC model, including the recent H.E.S.S. spectrum.
}
\label{fig-CenA_MultiBlobUpdated}
\end{figure}

Superimposing the recent H.E.S.S. spectrum to our early prediction for the VHE flux of Cen\,A obtained with the multi-blob SSC model (see Fig.~\ref{fig-CenA_MultiBlobUpdated}), even though being only one of the many possible interpretations, one can see that our prediction agrees quite well with the recent VHE data.

\section{Conclusions}

In conclusion, the detected VHE AGNs present a very heterogeneous broadband behavior and sometimes extreme flaring activities, especially in the case of blazars. The discovery of VHE $\gamma$-rays from one of the nearest AGN, Cen\,A, clearly establishes radio galaxies as a new class of VHE emitting sources.

With the fifth telescope of the H.E.S.S. experiment (the H.E.S.S.\,II project) coming online soon, and with the data from {\it Fermi}, the peak of the second component in the SEDs of blazars is expected to be detected, which will strongly constrain all the radiative models available for the processes involved in the emission from AGN jets.

With the advent of CTA (\v{C}erenkov Telescope Array\footnote{\url{http://www.cta-observatory.org/}}) and AGIS (Advanced Gamma-ray Imaging System\footnote{\url{http://www.agis-observatory.org/}}), many more discoveries are expected, with probably towards a thousand sources to be discovered.

\begin{acknowledgements}
I am very grateful to the organizers of the Frascati Workshop 2009 on ``Multifrequency Behaviour of High Energy Cosmic Sources'' (Vulcano, Italy) for their invitation to speak, and for creating a very pleasant and stimulating atmosphere for discussions during this workshop.

It is my pleasure to deeply thank my PhD advisors, Dr.~Catherine Boisson and Dr.~H\'el\`ene Sol, and my colleague Dr.~Andreas Zech from the LUTH for discussions and support during my PhD study. Finally, my thanks go to all the members of the H.E.S.S. collaboration for their technical support and for many stimulating discussions.
\end{acknowledgements}

\bibliographystyle{aa}

\begin{thebibliography}{23}
\expandafter\ifx\csname natexlab\endcsname\relax\def\natexlab#1{#1}\fi

\bibitem[{{Abdo}(2009)}]{2009arXiv0902.1559A}
{Abdo}, A.~A., {et~al.} ({\it Fermi}/LAT collaboration) 2009, \texttt{arXiv:0902.1559}

\bibitem[{{Aharonian} {et~al.}(2005){Aharonian}, {Akhperjanian}, {Bazer-Bachi},
  {Beilicke}, {Benbow}, {Berge}, {Bernl{\"o}hr}, {Boisson}, {Bolz}, {Borrel},
  {Braun}, {Breitling}, {Brown}, {Chadwick}, {Chounet}, {Cornils},
  {Costamante}, {Degrange}, {Dickinson}, {Djannati-Ata{\"i}}, {O'C.~Drury},
  {Dubus}, {Emmanoulopoulos}, {Espigat}, {Feinstein}, {Fontaine}, {Fuchs},
  {Funk}, {Gallant}, {Giebels}, {Gillessen}, {Glicenstein}, {Goret},
  {Hadjichristidis}, {Hauser}, {Heinzelmann}, {Henri}, {Hermann}, {Hinton},
  {Hofmann}, {Holleran}, {Horns}, {Jacholkowska}, {de Jager}, {Kh{\'e}lifi},
  {Komin}, {Konopelko}, {Latham}, {Le Gallou}, {Lemi{\`e}re},
  {Lemoine-Goumard}, {Leroy}, {Lohse}, {Martin}, {Martineau-Huynh},
  {Marcowith}, {Masterson}, {McComb}, {de Naurois}, {Nolan}, {Noutsos},
  {Orford}, {Osborne}, {Ouchrif}, {Panter}, {Pelletier}, {Pita},
  {P{\"u}hlhofer}, {Punch}, {Raubenheimer}, {Raue}, {Raux}, {Rayner}, {Reimer},
  {Reimer}, {Ripken}, {Rob}, {Rolland}, {Rowell}, {Sahakian}, {Saug{\'e}},
  {Schlenker}, {Schlickeiser}, {Schuster}, {Schwanke}, {Siewert}, {Sol},
  {Spangler}, {Steenkamp}, {Stegmann}, {Tavernet}, {Terrier}, {Th{\'e}oret},
  {Tluczykont}, {Vasileiadis}, {Venter}, {Vincent}, {V{\"o}lk}, \&
  {Wagner}}]{2005A+A...442..895A}
{Aharonian}, F.~A., {et~al.} (H.E.S.S. collaboration) 2005, \aap, 442, 895

\bibitem[{{Aharonian} {et~al.}(2006){Aharonian}, {Akhperjanian}, {Bazer-Bachi},
  {Beilicke}, {Benbow}, {Berge}, {Bernl{\"o}hr}, {Boisson}, {Bolz}, {Borrel},
  {Braun}, {Brown}, {B{\"u}hler}, {B{\"u}sching}, {Carrigan}, {Chadwick},
  {Chounet}, {Coignet}, {Cornils}, {Costamante}, {Degrange}, {Dickinson},
  {Djannati-Ata{\"i}}, {O'C.~Drury}, {Dubus}, {Egberts}, {Emmanoulopoulos},
  {Espigat}, {Feinstein}, {Ferrero}, {Fiasson}, {Fontaine}, {Funk}, {Funk},
  {F{\"u}{\ss}ling}, {Gallant}, {Giebels}, {Glicenstein}, {Goret},
  {Hadjichristidis}, {Hauser}, {Hauser}, {Heinzelmann}, {Henri}, {Hermann},
  {Hinton}, {Hoffmann}, {Hofmann}, {Holleran}, {Hoppe}, {Horns},
  {Jacholkowska}, {de Jager}, {Kendziorra}, {Kerschhaggl}, {Kh{\'e}lifi},
  {Komin}, {Konopelko}, {Kosack}, {Lamanna}, {Latham}, {Le Gallou},
  {Lemi{\`e}re}, {Lemoine-Goumard}, {Lenain}, {Lohse}, {Martin},
  {Martineau-Huynh}, {Marcowith}, {Masterson}, {Maurin}, {McComb}, {Moulin},
  {de Naurois}, {Nedbal}, {Nolan}, {Noutsos}, {Orford}, {Osborne}, {Ouchrif},
  {Panter}, {Pelletier}, {Pita}, {P{\"u}hlhofer}, {Punch}, {Ranchon},
  {Raubenheimer}, {Raue}, {Rayner}, {Reimer}, {Ripken}, {Rob}, {Rolland},
  {Rosier-Lees}, {Rowell}, {Sahakian}, {Santangelo}, {Saug{\'e}}, {Schlenker},
  {Schlickeiser}, {Schr{\"o}der}, {Schwanke}, {Schwarzburg}, {Schwemmer},
  {Shalchi}, {Sol}, {Spangler}, {Spanier}, {Steenkamp}, {Stegmann}, {Superina},
  {Tam}, {Tavernet}, {Terrier}, {Tluczykont}, {van Eldik}, {Vasileiadis},
  {Venter}, {Vialle}, {Vincent}, {V{\"o}lk}, {Wagner}, \&
  {Ward}}]{2006Sci...314.1424A}
{Aharonian}, F.~A., {et~al.} (H.E.S.S. collaboration) 2006, Science, 314, 1424

\bibitem[{{Aharonian} {et~al.}(2007){Aharonian}, {Akhperjanian}, {Bazer-Bachi},
  {Behera}, {Beilicke}, {Benbow}, {Berge}, {Bernl{\"o}hr}, {Boisson}, {Bolz},
  {Borrel}, {Boutelier}, {Braun}, {Brion}, {Brown}, {B{\"u}hler},
  {B{\"u}sching}, {Bulik}, {Carrigan}, {Chadwick}, {Clapson}, {Chounet},
  {Coignet}, {Cornils}, {Costamante}, {Degrange}, {Dickinson},
  {Djannati-Ata{\"i}}, {Domainko}, {Drury}, {Dubus}, {Dyks}, {Egberts},
  {Emmanoulopoulos}, {Espigat}, {Farnier}, {Feinstein}, {Fiasson},
  {F{\"o}rster}, {Fontaine}, {Funk}, {Funk}, {F{\"u}{\ss}ling}, {Gallant},
  {Giebels}, {Glicenstein}, {Gl{\"u}ck}, {Goret}, {Hadjichristidis}, {Hauser},
  {Hauser}, {Heinzelmann}, {Henri}, {Hermann}, {Hinton}, {Hoffmann}, {Hofmann},
  {Holleran}, {Hoppe}, {Horns}, {Jacholkowska}, {de Jager}, {Kendziorra},
  {Kerschhaggl}, {Kh{\'e}lifi}, {Komin}, {Kosack}, {Lamanna}, {Latham}, {Le
  Gallou}, {Lemi{\`e}re}, {Lemoine-Goumard}, {Lenain}, {Lohse}, {Martin},
  {Martineau-Huynh}, {Marcowith}, {Masterson}, {Maurin}, {McComb}, {Moderski},
  {Moulin}, {de Naurois}, {Nedbal}, {Nolan}, {Olive}, {Orford}, {Osborne},
  {Ostrowski}, {Panter}, {Pedaletti}, {Pelletier}, {Petrucci}, {Pita},
  {P{\"u}hlhofer}, {Punch}, {Ranchon}, {Raubenheimer}, {Raue}, {Rayner},
  {Renaud}, {Ripken}, {Rob}, {Rolland}, {Rosier-Lees}, {Rowell}, {Rudak},
  {Ruppel}, {Sahakian}, {Santangelo}, {Saug{\'e}}, {Schlenker}, {Schlickeiser},
  {Schr{\"o}der}, {Schwanke}, {Schwarzburg}, {Schwemmer}, {Shalchi}, {Sol},
  {Spangler}, {Stawarz}, {Steenkamp}, {Stegmann}, {Superina}, {Tam},
  {Tavernet}, {Terrier}, {van Eldik}, {Vasileiadis}, {Venter}, {Vialle},
  {Vincent}, {Vivier}, {V{\"o}lk}, {Volpe}, {Wagner}, {Ward}, \&
  {Zdziarski}}]{2007ApJ...664L..71A}
{Aharonian}, F.~A., {et~al.} (H.E.S.S. collaboration) 2007, \apjl, 664, L71

\bibitem[{{Aharonian} {et~al.}(2009){Aharonian}, {Akhperjanian}, {Anton}, {de
  Almeida}, {Bazer-Bachi}, {Becherini}, {Behera}, {Benbow}, {Bernl{\"o}hr},
  {Boisson}, {Bochow}, {Borrel}, {Brion}, {Brucker}, {Brun}, {B{\"u}hler},
  {Bulik}, {B{\"u}sching}, {Boutelier}, {Chadwick}, {Charbonnier}, {Chaves},
  {Cheesebrough}, {Chounet}, {Clapson}, {Coignet}, {Dalton}, {Daniel},
  {Davids}, {Degrange}, {Deil}, {Dickinson}, {Djannati-Ata{\"i}}, {Domainko},
  {Drury}, {Dubois}, {Dubus}, {Dyks}, {Dyrda}, {Egberts}, {Emmanoulopoulos},
  {Espigat}, {Farnier}, {Feinstein}, {Fiasson}, {F{\"o}rster}, {Fontaine},
  {F{\"u}{\ss}ling}, {Gabici}, {Gallant}, {G{\'e}rard}, {Giebels},
  {Glicenstein}, {Gl{\"u}ck}, {Goret}, {G{\"o}hring}, {Hauser}, {Hauser},
  {Heinz}, {Heinzelmann}, {Henri}, {Hermann}, {Hinton}, {Hoffmann}, {Hofmann},
  {Holleran}, {Hoppe}, {Horns}, {Jacholkowska}, {de Jager}, {Jahn}, {Jung},
  {Katarzy{\'n}ski}, {Katz}, {Kaufmann}, {Kendziorra}, {Kerschhaggl},
  {Khangulyan}, {Kh{\'e}lifi}, {Keogh}, {Klu{\'z}niak}, {Kneiske}, {Komin},
  {Kosack}, {Lamanna}, {Latham}, {Lenain}, {Lohse}, {Marandon}, {Martin},
  {Martineau-Huynh}, {Marcowith}, {Maurin}, {McComb}, {Medina}, {Moderski},
  {Moulin}, {Naumann-Godo}, {de Naurois}, {Nedbal}, {Nekrassov}, {Niemiec},
  {Nolan}, {Ohm}, {Olive}, {de O{\~n}a Wilhelmi}, {Orford}, {Ostrowski},
  {Panter}, {Arribas}, {Pedaletti}, {Pelletier}, {Petrucci}, {Pita},
  {P{\"u}hlhofer}, {Punch}, {Quirrenbach}, {Raubenheimer}, {Raue}, {Rayner},
  {Renaud}, {Rieger}, {Ripken}, {Rob}, {Rosier-Lees}, {Rowell}, {Rudak},
  {Rulten}, {Ruppel}, {Sahakian}, {Santangelo}, {Schlickeiser}, {Sch{\"o}ck},
  {Schr{\"o}der}, {Schwanke}, {Schwarzburg}, {Schwemmer}, {Shalchi}, {Sikora},
  {Skilton}, {Sol}, {Spangler}, {Stawarz}, {Steenkamp}, {Stegmann}, {Superina},
  {Szostek}, {Tam}, {Tavernet}, {Terrier}, {Tibolla}, {Tluczykont}, {van
  Eldik}, {Vasileiadis}, {Venter}, {Venter}, {Vialle}, {Vincent}, {Vink},
  {Vivier}, {V{\"o}lk}, {Volpe}, {Wagner}, {Ward}, {Zdziarski}, \&
  {Zech}}]{2009ApJ...695L..40A}
{Aharonian}, F.~A., {et~al.} (H.E.S.S. collaboration) 2009, \apjl, 695, L40

\bibitem[{{Aharonian}(2002)}]{2002MNRAS.332..215A}
{Aharonian}, F.~A. 2002, \mnras, 332, 215

\bibitem[{{Albert} {et~al.}(2007)}]{2007ApJ...669..862A} Albert, J., {et~al.} (MAGIC collaboration) 2007, \apj, 669, 862 

\bibitem[Chadwick et al.(1999)]{1999ApJ...513..161C} Chadwick, P.~M., {et~al.} 1999, \apj, 513, 161 

\bibitem[{{Croston} {et~al.}(2009){Croston}, {Kraft}, {Hardcastle},
  {Birkinshaw}, {Worrall}, {Nulsen}, {Penna}, {Sivakoff}, {Jord{\'a}n},
  {Brassington}, {Evans}, {Forman}, {Gilfanov}, {Goodger}, {Harris}, {Jones},
  {Juett}, {Murray}, {Raychaudhury}, {Sarazin}, {Voss}, \&
  {Woodley}}]{2009MNRAS.395.1999C}
{Croston}, J.~H., {Kraft}, R.~P., {Hardcastle}, M.~J., {et~al.} 2009, \mnras,
  395, 1999

\bibitem[{{Ghisellini} {et~al.}(2005){Ghisellini}, {Tavecchio}, \&
  {Chiaberge}}]{2005A+A...432..401G}
{Ghisellini}, G., {Tavecchio}, F., \& {Chiaberge}, M. 2005, \aap, 432, 401

\bibitem[{{Katarzy{\'n}ski} {et~al.}(2008){Katarzy{\'n}ski}, {Lenain}, {Zech},
  {Boisson}, \& {Sol}}]{2008MNRAS.390..371K}
{Katarzy{\'n}ski}, K., {Lenain}, J.-P., {Zech}, A., {Boisson}, C., \& {Sol}, H.
  2008, \mnras, 390, 371

\bibitem[{{Katarzy{\'n}ski} {et~al.}(2003){Katarzy{\'n}ski}, {Sol}, \&
  {Kus}}]{2003A+A...410..101K}
{Katarzy{\'n}ski}, K., {Sol}, H., \& {Kus}, A. 2003, \aap, 410, 101

\bibitem[{{Krawczynski} {et~al.}(2004){Krawczynski}, {Hughes}, {Horan},
  {Aharonian}, {Aller}, {Aller}, {Boltwood}, {Buckley}, {Coppi}, {Fossati},
  {G{\"o}tting}, {Holder}, {Horns}, {Kurtanidze}, {Marscher}, {Nikolashvili},
  {Remillard}, {Sadun}, \& {Schr{\"o}der}}]{2004ApJ...601..151K}
{Krawczynski}, H., {Hughes}, S.~B., {Horan}, D., {et~al.} 2004, \apj, 601, 151

\bibitem[{{Lenain} {et~al.}(2008{\natexlab{a}}){Lenain}, {Benbow}, {Boisson},
  {B{\"u}hler}, {Costamante}, {Giebels}, {Katarzy{\'n}ski}, {Pita}, {Punch},
  {Raue}, {Sol}, {Ne}, {Superina}, {Volpe}, \& {Zech}}]{2008AIPC.1085..415L}
{Lenain}, J.-P., {et~al.} (for the H.E.S.S. collaboration) 2008{\natexlab{a}}, in
  American Institute of Physics Conference Series, Vol. 1085, 415--418

\bibitem[{{Lenain} {et~al.}(2008{\natexlab{b}}){Lenain}, {Boisson}, {Sol}, \&
  {Katarzy{\'n}ski}}]{2008A+A...478..111L}
{Lenain}, J.-P., {Boisson}, C., {Sol}, H., \& {Katarzy{\'n}ski}, K.
  2008{\natexlab{b}}, \aap, 478, 111

\bibitem[{{Mannheim} {et~al.}(1991){Mannheim}, {Biermann}, \&
  {Kruells}}]{1991A+A...251..723M}
{Mannheim}, K., {Biermann}, P.~L., \& {Kruells}, W.~M. 1991, \aap, 251, 723

\bibitem[{{Maraschi} {et~al.}(1992){Maraschi}, {Ghisellini}, \&
  {Celotti}}]{1992ApJ...397L...5M}
{Maraschi}, L., {Ghisellini}, G., \& {Celotti}, A. 1992, \apjl, 397, L5

\bibitem[{{McKinney}(2006)}]{2006MNRAS.368.1561M}
{McKinney}, J.~C. 2006, \mnras, 368, 1561

\bibitem[{{M{\"u}cke} {et~al.}(2003){M{\"u}cke}, {Protheroe}, {Engel},
  {Rachen}, \& {Stanev}}]{2003APh....18..593M}
{M{\"u}cke}, A., {Protheroe}, R.~J., {Engel}, R., {Rachen}, J.~P., \& {Stanev},
  T. 2003, Astroparticle Physics, 18, 593

\bibitem[{{Neronov} \& {Aharonian}(2007)}]{2007ApJ...671...85N}
{Neronov}, A. \& {Aharonian}, F.~A. 2007, \apj, 671, 85

\bibitem[{{Orellana} \& {Romero}(2009)}]{2009arXiv0902.0731O}
{Orellana}, M. \& {Romero}, G.~E. 2009, \texttt{arXiv:0902.0731}

\bibitem[{{Reimer} {et~al.}(2004){Reimer}, {Protheroe}, \&
  {Donea}}]{2004A+A...419...89R}
{Reimer}, A., {Protheroe}, R.~J., \& {Donea}, A.-C. 2004, \aap, 419, 89

\bibitem[{{Rieger} \& {Aharonian}(2008)}]{2008A+A...479L...5R}
{Rieger}, F.~M. \& {Aharonian}, F.~A. 2008, \aap, 479, L5

\bibitem[{{Sikora} {et~al.}(1994){Sikora}, {Begelman}, \&
  {Rees}}]{1994ApJ...421..153S}
{Sikora}, M., {Begelman}, M.~C., \& {Rees}, M.~J. 1994, \apj, 421, 153

\bibitem[{{Stawarz} {et~al.}(2006){Stawarz}, {Aharonian}, {Wagner}, \&
  {Ostrowski}}]{2006MNRAS.371.1705S}
{Stawarz}, {\L}., {Aharonian}, F., {Wagner}, S., \& {Ostrowski}, M. 2006,
  \mnras, 371, 1705

\end{thebibliography}

\bigskip
\bigskip
\noindent {\bf DISCUSSION}

\bigskip
\noindent {\bf JAMES H. BEALL:} In your multi-blob model, do you imagine them emerging sequentially or in different directions simultaneously~?

\bigskip
\noindent {\bf JEAN-PHILIPPE LENAIN:} It is a stationary model, thus the dynamics of the bulk motion of the blobs is not considered.

Moreover, it should just be seen as an illustrative sketch, as implemented in our code. Physically, one should consider it as representing an inhomogeneous flow continuously crossing a stationary shock front within the jet, just beyond the Alfv\'en surface.

\end{document}